%Paper: hep-th/9207107
%From: Kresimir Demeterfi <ST402758@brownvm.brown.edu>
%Date: Thu, 30 Jul 92 14:47:10 EDT

%%%%%%%%%%%%%%%%%%%%   Please use PHYZZX   %%%%%%%%%%%%%%%%%%%%%%%%%%%

\hsize=6.0in
\vsize=8.9in
\hoffset=0.0in
\voffset=0.0in

\def\IR{\relax{\rm I\kern-.18em R}}
\def\IN{\relax{\rm I\kern-.18em N}}
\def\IC{\relax\thinspace\hbox{$\inbar\kern-.3em{\rm C}$}}
\def\IZ{\relax\ifmmode\mathchoice
   {\hbox{\cmss Z\kern-.4em Z}}{\hbox{\cmss Z\kern-.4em Z}}
   {\lower.9pt\hbox{\cmsss Z\kern-.4em Z}}
   {\lower1.2pt\hbox{\cmsss Z\kern-.4em Z}}\else{\cmss Z\kern-.4em Z}\fi}

\FRONTPAGE
\line{\hfill BROWN-HET-855}
\line{\hfill March 1992}
\vskip1.5truein
\titlestyle{{$w_{\infty}$-Currents in 3-Dimensional Toda Theory
}}
\bigskip
\author{Jean AVAN\foot{On leave of absence from L.P.T.H.E. Paris 6
(CNRS UA 280, France).  Work sponsored by Brown University Exchange
Program  P. I. 135.} }
\centerline{{\it Department of Physics}}
\centerline{{\it Brown University, Providence, RI 02912, USA}}
\bigskip
\abstract

\noindent Chiral densities obeying a $w_{\infty}$ Poisson--bracket
algebra are constructed for the $2+1\,\, A_{\infty}$ -- Toda field
theory, using its alternative $w_{\infty}$ -- Toda
representation.  They are obtained from formal traces of powers
of the Lax operator.  The spin 2 and 3 currents are explicitely
derived, and the consistency of their Poisson algebra is checked.

\endpage

\noindent {\bf 1. \underbar{Introduction}}

The construction and study of extended conformal symmetry algebras,
specifically $W_N$-algebras, has been a major development in a number
of domains of classical and quantum field theory [1].  These algebras
are generically non--linear for $N>2$ $(W_2$ is the Virasoro algebra),
but the notion of a linearizing limit $N\rightarrow \infty$ was shown to be
mean
although not unique [2].  This gave rise to a number of (globally
denoted) $W_{\infty}$ algebras, the simplest one being isomorphic to the
algebra of 2-dimensional area--preserving diffeomorphisms of the
plane $\IR^2$ $Sdiff(2)$, also noted $w_{\infty}$ [3].   Viewed as a classical
algebra of functions over a 2--dimensional phase space, $w_{\infty}$
admits a quantum linear deformation $W_{\infty}$; the general
consistent linear deformations, also denoted $W_{\infty}$, were constructed
in [4].  $w_{\infty}$ is defined by the Lie bracket:
$$\bigl[ W_{m_1}^{n_1} \, \, , W_{m_2}^{n_2} \bigr] = \bigl( (m_2 -1
) n_1 - (m_1 - 1) n_2 \bigr) \, W_{m_1 + m_2 -2}^{n_1 + n_2} \quad m \geq
2\eqno\eq$$
The ``classical" $w_{\infty}$ and ``quantum" $W_{\infty}$ algebras were
realized using 2--dimensional bosonic [1,6] and fermionic [5] fields,
and it was suggested that a representation in term of chiral currents
from a 3--dimensional classical field should exist for $w_{\infty}$ [2].
The argument runs as follows:

The non--linear $W_N$--algebras are naturally associated to the classical
$sl(N+1)$ Toda 2--dimensional field theory: starting from the Lax
system in the fundamental representation of $sl (N+1)$, the
$W_N$ generators are obtained from
the coefficients (chiral densities) of the  $N+1$--order differential
equation obeyed by the projection of the Lax system solution on a
highest--weight vector of $sl(N+1)$ [7].  The relation between the
initial canonical variables of the $sl(N)$
Toda field theory and the generators $\{ u_n\}$ of $W_{N-1}$ is in
fact a generalized Miura transformation for the corresponding KdV--
type hierarchy [7-9].  The $W_N$--algebra is thus  related
to the second Gelfand--Dikii Poisson structure of the $N$--th KdV
hierarchy [8].  The $w_{\infty}$ algebra is then obtained rigorously
by taking a consistently defined $N\rightarrow \infty$ limit of the second
Gelfand-Dikii bracket for the $N$--th KdV hierarchy [2], and redefining the
generators $w_n$ as the limit $(N\rightarrow\infty )$ of the
densities generating the conserved
Hamiltonians of the dispersionless KdV hierarchy [18].
Hence it is to be expected that the
Toda--$sl (\infty )$ field theory, once it is rigorously defined,
provide us with a realization of $w_{\infty}$--algebra in term of
chiral ``Hamiltonian" densities.  Such a construction was indeed
suggested in [2], using the notion of continuous Lie algebra [10]
(specifically here a continuous $A_{\infty}$ algebra) to define the $2 +
1$ dimensional Toda theory as a natural candidate to generate a
representation of $w_{\infty}$.  However the explicit construction of
the generators of $w_{\infty}$ was not achieved in this framework
due to the difficulty
in defining the $N\rightarrow \infty$ limit of the chiral currents for
the $N$--th KdV hierarchy.

We wish to present here an explicit derivation of the classical $w_{\infty}$--
Poisson algebra from the 2+1--dimensional field of the Toda theory
considered in [2].  Rather than trying to define the limit
$N\rightarrow\infty$ of the Miura--transformed generators of $W_N$,
however, we shall directly use the recently introduced [11] Lax
representation of the $2+1$ dimensional Toda
as a $w_{\infty}$--Toda field theory -- instead of a $sl (\infty )$
Toda field theory -- and use the solution ${\cal L} (z, \bar{z} )$ of
this Lax system to define the relevant generators.  Our major goal is
the explicit derivation of the yet unknown [2] spin--2 current,
hereafter denoted
$w_3$.  Indeed, in a $w_{\infty}$--algebra, as indicated in [2], it is
sufficient to know $w_2$ and $w_3$ to reconstruct the full
algebra by Poisson brackets.

\noindent{\bf 2.  \underbar{The 2+1--Dimensional Toda Hierarchy}}

The Toda $sl(N+1)$ 2--dimensional field theory is described by the
equations
$$\partial\bar{\partial} \varphi_i = {1\over 4} \, \sum_{\{ \alpha \}}
\quad \alpha_i \, \exp (\alpha \cdot\varphi )\eqno\eq$$
where $\alpha$ are the simple roots of $sl (N+1)$.  It is obtained as
a zero--curvature condition from a Lax pair [20-22]:
$$\eqalign{ \partial_x & + \Bigl( {1\over 2} \, \varphi,_t \cdot H +
{1\over 2} \, \sum_{\{ \alpha \}} \, \exp \bigl( {1\over 2} \, \alpha
\cdot \varphi\bigr) \, \bigl( E_{\alpha} + E_{-\alpha}\bigr)\Bigr) \cdot
\Psi = 0\cr
\partial_t & + \Bigl( {1\over 2} \, \varphi ,_x \cdot H + {1\over 2}
\, \sum_{\{\alpha\}} \, \exp \bigl( {1\over 2} \alpha \cdot \varphi \bigr)
\, \bigl( E_{\alpha} - E_{-\alpha}\bigr)\Bigr) \cdot \Psi = 0}\eqno\eq$$
or, in light--cone coordinates $\bigl( z , \bar{z} = x \pm t\bigr)$:
$$\eqalign{\bigl( \partial + \partial \Phi + \exp \, ad\Phi \cdot
E^+ \bigr) T & = 0\cr
\bigl( \bar{\partial} - \bar{\partial}\Phi + \exp - ad\Phi
\cdot E^-) T & = 0 }$$
which is gauge--equivalent to
$$\eqalign{ \bigl( \partial + 2 \partial \Phi + E^+ \bigr)
\tilde{T} & = 0\cr
\bigl( \bar{\partial} + \exp - 2 ad\Phi \cdot E^- \bigr)
\tilde{T} & = 0 } \eqno\eq$$

Here $\Phi = \varphi \cdot H = \sum \varphi_i h_i$, $\{ h_i \}$ being
the Cartan subalgebra of $sl (N+1)$ and $E^+ , E^-$ are
the sum of (respectively) positive and negative simple roots. This
light--cone representation
relates the $sl(n)$--Toda theory to a constrained $sl (n) - WZW$ model, as
emphasized in [9,12].  Note that the Poisson structure of the dynamical
variables, which was initially ultralocal in (3):
$$\{ \varphi^i,_t = \Pi^i , \varphi^j \} \cong \delta (x-y)
K^{ij}\eqno\eq$$
has now a more natural, non--ultralocal form in (4):
$$\{ \partial \varphi ^i , \, \partial \varphi ^j \} \simeq \delta '
(x-y) K^{ij}$$
$K^{ij}$ being the Cartan matrix of $sl (N+1)$.  This change of
``fundamental" dynamical variables is an important feature in the
shift from $sl (\infty )$
to $w_{\infty}$ as a fundamental underlying Lie algebra, since a
$\delta '$--Poisson algebra naturally leads to a $w_{\infty}$--
algebra [1].

The limit $N\rightarrow \infty$ of (2) is defined by taking [13]:
$$u (z, \bar{z} , s = s_0 + i\Delta ) \equiv  \Delta \varphi_i (z ,\bar{z}
)\quad ; \quad \Delta = \, {s_1 - s_0\over N}\eqno\eq$$
$[s_0 , s_1 ]$ is the arbitrary interval of variation for the third continuous
v
(2) then becomes the ``2+1 Toda equation":
$$\partial\bar{\partial} u = - \partial_s \, \exp \, \partial_s
u\eqno\eq$$
There exists another Lax representation of (7) using directly a
$w_{\infty}$--algebra structure instead of considering the limit
$N\rightarrow \infty$ of the original $sl (N)$ structure in (3).
One introduces two (initially independent) Lax operators:
$$\eqalign{ {\cal L} (\lambda , s) & = \lambda + u_0 + \sum_{n\geq 1} \,
\lambda^{-n} \, u_n (s)\cr
\hat{\cal L} (\lambda , s) & = \sum_{n\geq 1} \, \lambda^n \, \hat{u}_n (s)
}\eqno\eq$$
The 3-d Toda hierarchy, including as a particular example (7), is
written as:
$$\eqalign{\bar{\partial}_{(n)} {\cal L} & = \{ \bar{B}_{(n)} , {\cal
L}\}_{S{\r
\partial_{(n)} ({\cal L}) = \{ B_{(n)} , {\cal L}\}_{S{\rm diff}}\cr
\bar{\partial}_{(n)} \hat{{\cal L}}& = \{ \bar{B}_{(n)} , \hat{{\cal
L}} \} S_{{\rm diff}} ; \partial_{(n)} \hat{{\cal L}} = \{ B_{(n)} , \hat{{\cal
L}} \} S_{{\rm diff}} } \eqno\eq$$
where:
$$\eqalign{ \bar{B}_{(n)} & \equiv \bigl( \hat{{\cal L}}^{-n} \bigr)  \leq -
1 (i.e.\,\, {\rm elements \,\, of\,\, order\,\, in\,\,\lambda \,\, smaller\,\,
than\,\, -1})\cr
B_{(n)} & = \bigl( {\cal L}^n \bigr) \geq 0 }\eqno\eq$$
$$\{ F,G\}_{S{\rm diff}} \equiv \, \lambda \, {\partial F\over
\partial\lambda} \, {\partial G\over \partial s} \, - \, \lambda \,
{\partial F\over \partial s} \, {\partial G\over \partial\lambda}
\eqno\eq$$
(7) will be obtained from the Frobenius compatibility condition of (9)
between $\bar{B}_{(1)}$ and $B_{(1)}$.  This construction has a very
straightforward interpretation in term of a {\it two}--dimensional $w_{\infty}$
Toda field theory.  (11) is a Poisson bracket which is dual to  the
Lie algebra structure of $w_{\infty} \simeq \, Sdiff$ (2) and can
thus be understood as a
Lie bracket acting on a functional representation of $w_{\infty}$.
 Indeed there exists a Lie algebra isomorphism between
$w_{\infty}$ and the space $\{ F (\lambda , s)\}$ of doubly analytic
functions with the Lie bracket (11). (see for instance [17] for an example
of this identification).   The Hamiltonians (10) are
canonical Toda--type hamiltonians:  for instance the $z$--evolution is
described by the projection of ${\cal L}^n$ on the positive Borel subalgebra,
rewritten as:
$$\bigl( {\cal L}^n \bigr)_{\geq 0} = R. dh({\cal L})\eqno\eq$$
where  $h({\cal L}) = Tr \, {\cal L}^{n+1}$ is a natural adjoint invariant in
$w_{\infty}$ and $R$ is the $r$-matrix operator $P_+ - P_- $
describing the splitting of $w_{\infty}$ into
positive and negative powers of $\lambda$. (see [14], [17] and [20,22])

This interpretation also helps understanding straightforwardly why
the 3d Toda theory has an underlying $w_{\infty}$--algebra of dressing
transformations [13].  It is a consequence of it being in fact
a two-dimensional $w_{\infty}$ Toda theory.  This algebra indeed appears very
naturally in [11] as a consequence of the Riemann-Hilbert
representation of the Lax system (9) (see also [15] for general
properties of dressing transformations in an algebraic framework).

Finally we wish to comment on the relation uncovered in this way
between $sl (\infty)$ and $w_{\infty}$.  The 3d Toda theory can
alternatively be described by an $sl (\infty)$--algebra structure (3)
or a $w_{\infty}$--algebra structure (9).  The links between the two
are provided by the light--cone rotation (4), the gauge transformation
$T \rightarrow \exp - ad\Phi \cdot T$ in the zero--curvature condition for
(4), and the definition of the continuous limit which ultimately
transform (4) into a group--version of the algebra--valued Lax system
(9).  It is known that $sl
(\infty)$ is isomorphic to $W_{1 +\infty}$ as can be seen from
their respective bilinear free fermion representations [5] (see also
[24]).  Hence $w_{\infty}$
now appears as a traceless, non--linearly embedded subalgebra of
$W_{1+\infty}$. We hope to develop on this soon [16].

\noindent{\bf 3.  \underbar{The Chiral $w_3$ Current}}

The $w_{\infty}$--algebra representation (9) of the 3-d Toda hierarchy
leads to formally $\partial$--and $\bar{\partial}$ conserved densities defined
a
$$w_n = Tr \, {\cal L}^n\eqno\eq$$
where $Tr$ is the $w_{\infty}$--adjoint invariant trace, represented as $\oint
\, {d\lambda\over\lambda} \wedge ds$,  once $w_{\infty}$ is
represented by the functional bracket (11).  One must however be very
careful when using the adjoint--invariance properties of this trace,
namely partial integration with respect to $\lambda$ and $s$.  Indeed
the transition from the discrete $i$--index to the continuous
$s$--variable does not insure that $u_n$ or $\hat{u}_n$ are analytic
in $s$.  Hence ${\cal L}$ does not necessarily represent an element of the
$w_{\infty}$--algebra, viewed as the set of analytic functions in
$\lambda$ and $s$ with the bracket (10), but rather lives in
an extension of
$w_{\infty}$.  The $\partial$ and
$\bar{\partial}$--``conservation" of $Tr {\cal L}^n$ in fact is
realized up to
integrals of total derivatives $\oint ds \partial_s F$, but with a
priori non--analytic functions $F$, and is therefore not exact.
However, these formal conservation laws can be turned into exact
conservation laws of analytic current densities by suitable
redefinitions of these current densities $w_n$,
which we shall soon describe.

{}From (8) and (9), considering only the evolution under $z$ and
$\bar{z}$ triggered by $\bar{B}_1$ and $B_1$, we have:
$$\eqalign{ &\partial u_0 = \partial_s u_1
\cr
&\partial u_1 = \partial_s u_2 + u_1 \partial_s u_0
\cr
& \partial u_2 = \partial_s u_3 + 2 u_2 \partial_s u_0\cr}
\qquad
\eqalign{ \bar{\partial} u_0 =  & -
\partial_s \hat{V}_1 \, (\hat{V}_1 \equiv \hat{u}_1 ^{-1} )\cr
\bar{\partial} u_1 = & - \hat{V}_1 \partial_s u_0\cr
\bar{\partial} u_2 = & - \hat{V}_1 \partial_s u_1 + u_1 \partial_s \hat{V}_1
}\eqno\eq$$
$$\partial\hat{u}_1 = - \partial_s u_0 \cdot \hat{V}_1 \quad ({\rm
from \,\, (9) \,\, applied\,\, to\,\, \hat{{\cal L}})\,\,,  as\,\, it\,\,
must\,\, be} )\eqno\eq$$
Since we are dealing with a gauge--transformed $sl (\infty )$-Toda
theory, it remains consistent to impose $Tr {\cal L} = 0$, hence
$$\eqalign {\oint ds \cdot u_0 & = 0\cr
\Rightarrow \oint ds \cdot \partial_s u_1 & = 0 }\eqno\eq$$
We assume the minimal analyticity requirement, namely the analyticity
of $u_0$. From (16) if follows that $u_0$ has no $s^{-1}$--term,
hence from (14) $u_1$ is also analytic in $s$ which is indeed consistent with
(16).  However, nothing constrains
$u_{n\geq 2}$ to be analytic in $s$.

If we now set [11]:
$$\eqalign{ u_1 & = - \partial^2 \phi\cr
u_0 & = - \partial\partial_s \phi\cr
\hat{u}_1 & = \exp - \partial_s^2 \phi \quad ; \quad \phi\,\, {\rm
analytic \,\, in} \, s }\eqno\eq$$
thereby fulfilling as ``constraints" the $\partial$--equations, the
$\bar{\partial}$--``evolution" equation becomes:
$$\eqalign { & \partial\bar{\partial} \cdot \partial_s \phi  = -
\partial_s \cdot \exp - \partial_s^2 \phi\cr
& \quad\quad\Leftrightarrow \partial\bar{\partial} u  =
\partial_s \cdot \exp \,
\partial_s u\,\, {\rm for} \, \, u = \partial_s \phi\cr
&\quad\quad \Leftrightarrow\,  {\rm equation}  (7) }$$

\noindent Hence the Toda equation (7) is indeed a part of the Toda
hierarchy (8)-(9).  In fact, we shall from now on consider $\phi (z,
\bar{z}, s)$ as the relevant physical field.  It follows from (17)
and (14) that $\phi$
obeys $\partial_s (\partial\bar{\partial} \phi + \exp - \partial_s^2
\phi ) = 0$; a redefinition of $\phi$ can always take into account the
integration constant, and be chosen so as to obey:
$$\partial\bar{\partial} \phi + \exp - \partial_s ^2 \phi = 0 \eqno\eq$$
This equation will from now on be taken as the definition of the 2+1 Toda
field dynamics.  It implies the original Toda hierarchy once (17) is
implemented, hence ${\cal L}$ and $\hat{{\cal L}}$ obey the equations
(8)--(9).

Although the quantities $Tr {\cal L}^n$ are in principle not conserved
($\partial$ or $\bar{\partial}$--invariant) because of the
non--analyticity of all but two coefficients in ${\cal L}$, we shall however
derive exactly (not formally) $\bar{\partial}$--invariant new
densities, from such expressions.
These new densities, depending only on the analytic physical field
$\phi (z, \bar{z}, s)$, will be natural candidates to be identified
with the large-$N$ limit of the
chiral quantities $\{ w_n , n = 1\cdots N\}$ defining the $W_{N-1}$
algebra from a $sl(N)$ Toda field theory.

We recall indeed that the algebra $w_{\infty}$ was originally constructed in
[2] by considering the $N\rightarrow  \infty$ limit of the conserved
Hamiltonians ${\cal L}^{n/N}$ for a $N$--order KdV hierarchy, which is
in fact a gauge--transform of a $sl(N)$--Toda theory, as indicated in
[7,9].  This means that the generators of the linear $w_{\infty}$
algebra are in fact non--trivial polynomials in the $u_n$ Gelfand--
Dikii components, except the first two generators $w_2 = u_2$ and $w_3 = u_3$.
therefore no ambiguity in identifying the first 3 (or non--trivial 2)
chiral densities; and these are the only generators needed to
construct the full $w_{\infty}$ algebra.\nextline
The first three traces read:
$$\eqalign {w_1 & =  \int ds \, u_0 = 0\cr
{1\over 2} \, w_2 & = \int ds \, \bigl( {u_0^2\over 2} + u_1 \bigr) =
\int ds \bigl( {(\partial\partial_s \phi )^2\over 2} - \partial^2 \phi
\bigr)\cr
{1\over 3} \, w_3 & = \int ds \, \bigl( {u_0 ^3\over 3} + 2 u_0 u_1 +
u_2 ) = \int ds \bigl( {(\partial\partial_s\phi )^3\over 3} + 2
\partial\partial_s\phi \cdot \partial^2 \phi + u_2 \bigr) }\eqno\eq$$

$w_2$ turns out to be the improved energy--momentum tensor for
the Toda theory; it was already derived directly from the Lagrangian
of 3d Toda in [2].  $w_3$ is expressed in term of the a-priori
non--analytic function $u_2$.  However, a formal part--integration
converts $\int sds\cdot u_2 \,$ into $  - \int s \, \partial_s u_2 \cdot
ds = - \int sds (\partial u_1 - u_1 \partial_s u_0)$ (from (14))
which is an analytic density.
This ``partially--integrated" $w_3$ is now a well-defined function of
$z$ and $\bar{z}$.  It reads:
$$w_3 (z,\bar{z}) = \int ds \bigl\{ {(\partial\partial_s \phi )^3\over 3} + 2
(\partial\partial_s \phi ) \partial^2 \phi + s ( \partial^3 \phi +
\partial^2 \phi \,\partial\partial_s^2 \phi )\bigr\}\eqno\eq$$
Moreover, first without partial integration, one gets from (19) and
(14):
$$\eqalign {\bar{\partial} w_3 & = \int ds \cdot - u_0 ^2 \, \partial_s
\hat{V}_1 - 2 \, u_0 \partial_s u_0 \hat{V}_1 - 2 \, \partial_s
\hat{V}_1 u_1 + u_1 \partial_s \hat{V}_1 - \hat{V}_1 \partial_s u_1\cr
& = \int ds \cdot \partial_s \bigl( - u_0 ^2 \hat{V}_1 - u_1 \hat{V}_1
\bigr) = 0 \,\,{\rm since}\,\, u_0 , \hat{V}_1 \,\,{\rm and} \,\, u_1
\,\, {\rm are \,\, analytic} }\eqno\eq$$
Replacing, as one should, $u_2$ in (19) by its correctly defined
analytic partial--
integrated counterpart $\int s \partial_s u_2 ds$ does
not
modify this computation since this substitution can consistently
be undone once it is
noticed that $\bar{\partial} \, u_2$ gives a purely analytic
contribution to $\bar{\partial} w_3 $; thus the partial integration of
the $\bar{\partial}$--derived density is a consistent well-defined
procedure, and the analytic current $w_3$ $(z, \bar{z})$ defined by (20) is
exactly chiral.  In fact, $\bar{\partial} w_3 = 0$ can be obtained
directly as a consequence of the Toda equation (18).

A partial check of formula (20) is available.
General formulas for $w_n$--generators in a Toda $sl (N+1)$ theory are
quite cumbersome [7]; however the leading term of $w_3$ can quite easily be
computed, and turns out to be $\sum (\phi_i )^3$, leading directly
and unambiguously to $\int (\partial\partial_s \phi )^3 ds$ in the continuum.
This is indeed the leading order term of $w_3$ in (20).  In fact, this
property of $w_3 (z, \bar{z} )$ plus its chiral invariance should be enough to
completely fix its form to be (20).

To confirm that $w_3 (z)$ is the
correct
generator of the 3d--Toda $w_{\infty}$ algebra, we must now compute the
Poisson bracket  $\{ w_2 , w_3 \}$.  The Poisson structure for the 3d
Toda system follows from (5),(6) and (17):
$$\{ \partial\partial_s \phi (s,z) \, , \, \partial\partial_s \phi (s'
, z' )\} = \delta ' (z-z') \, \delta (s-s')\eqno\eq$$
We are here considering a light--cone picture where $z$ is a ``space"
variable and $\bar{z}$ a ``time" variable; the $PB$ then becomes a
function of $z$ and $s$ instead of $x$ and $s$.

The Poisson structure induced by (22) for $w_2$ leads directly
to a Virasoro algebra as described in [2,7]:
$$\{ w_2 (z) , w_2 (z') \} = \bigl( w_2 (z) + w_2 (z') \bigr)\, \delta
' (z - z ' )\eqno\eq$$
The Poisson bracket $\{ w_2 , w_3 \}$ is less straightforward to compute;
the previous formal partial integration of $u_2$ yielding the
definite form (20), is crucial in order to
rewrite this $PB$ under a closed form, using only (22).

Moreover, the form of $w_2$ and $w_3$ is such that consistent (not
formal!) partial integrations
can always be used to reduce Poisson brackets to a general factorized
form like (23).  The output, regrouping the 17 different terms
generated by this Poisson bracket, and using the $\delta$--
distribution properties:
$$\eqalign{ f (u_1) f(u_2 ) \delta ' (u_1 - u_2 ) & = {1\over 2}
\bigl( f^2 (u_1 ) + f^2 (u_2 )\bigr) \delta ' (u_1 - u_2 )\cr
\bigl( f (u_1 ) + f (u_2 )\bigr) \delta ' (u_1 - u_2 ) & = 2 f (u_2 ) \delta
' (u_1 - u_2 ) - f ' (u_1 ) \delta (u_1 - u_2 ) }\eqno\eq$$
is the characteristic $w_{\infty}$--algebra relation:
$$\{w_2 (z) , w_3 (z') \} = \bigl( 2 w_3 (z) + w_3 (z') \bigr) \,
\delta ' (z-z')\eqno\eq$$

Since $w_3 (z)$ is $\bar{\partial}$--invariant due to the Toda
equations, and obeys the correct Poisson relation with $w_2 (z)$, it
is perfectly consistent to interpret it as the spin--2 generator of the
$w_{\infty}$--algebra which was conjectured but not computed in [2].
Moreover, as we have seen, $w_3$ has the correct leading--order
behaviour of the $N\rightarrow \infty$ limit of the exact
$w_3$--density in
$sl(N)$ Toda.  Again we emphasize that $w_3 (z)$ defined in (20) is a
well--defined, exactly chiral current, and has exact Poisson brackets
of a spin--3 current (25).  In principle the full $w_{\infty}$--
algebra now follows from knowing $w_2$ (3) and $w_3$ (3).

All higher generators are in principle obtained from successive
Poisson brackets of
$w_3$ with itself.  Although the computations grow rapidly to an
unreasonable size, we shall explicitely derive the next generator
$w_4 (z)$, since it shows another interesting and tractable example
of how the formal density $Tr {\cal L}^n$ can be interpreted in a
suitable way to become a well--defined chiral density.  This moreover
provides a very non--trivial check on the validity of our
identification of $w_3$ with the spin--2 current.

\noindent{\bf 4.  \underbar{Higher generators}}

The computation of $\{ w_3 (z) , w_3 (z ')\}$ is rather long,
containing some 106 different terms multiplying distributions
$\delta ''' , \delta '' , \delta '$ and $\delta$.  It ultimately boils
down to:
$$\{ w_3 (z) , w_3 (z')\} = 2 \bigl( w_4 (z) + w_4 (z' ) \bigr) \delta
' (z - z' )\eqno\eq$$
$$\eqalign {& w_4 (z) = \int ds \bigl\{ {(\partial\partial_s\phi
)^4\over 4} \, - \, (\partial\partial_s \phi )^2 \partial ^2 \phi +
\partial^2 \partial_s \phi \cdot \partial\partial_s \phi\cdot
\partial\phi + s \cdot \partial^2 \partial_s \phi \cdot\cr
& - \partial^3 \phi \cdot \partial\phi + (\partial ^2 \phi )^2 - s \,
\partial^3 \phi \cdot \partial\partial_s \phi + {s^2\over 2} \bigl(
(\partial^2 \partial_s \phi )^2 + (\partial ^3 \partial_s \phi )
\partial \partial_s \phi \bigr) - {s^2\over 2} \, \partial^4 \phi \}
}\eqno\eq$$

Interestingly, this expression can be obtained from the formal chiral
density $Tr \, {\cal L}^4$ defined to be
$${1\over 4} Tr \, {\cal L}^4 = \int \bigl( {u_0^4\over 4} + 3 u_0 ^2
u_1 + {3\over 2} \, u_1^2 + 3 u_0 u_2 + u_3 \bigr) ds\eqno\eq$$
Partial--integration of $u_0 u_2 = - \partial\partial_s \phi \cdot
u_2$ as $\partial \phi \cdot \partial_s u_2$, then partial integration
of $u_3$ as $-s\partial_s u_3$ and furtherly $= {s^2\over 2}\,
\partial_s ^2 u_3$, transforms the ill--defined non--analytic
expression (28) into the analytic well-defined expression (27), using
the equations (14).

The whole scheme consists in formally partial--integrating the
ill--defined expressions involving $u_2$  and $u_3$ so as to get
``equivalent" but well--defined analytic expressions involving only the
field $\phi (z, \bar{z}, s)$ and its derivatives, using the hierarchy
equations (14).  In fact, we interpret (27), not (28) as the correct physical
density described by ``$Tr \, {\cal L}^4 $"; and similarly (20), not
(19), to be $Tr \, {\cal L}^3$.

The exact $\bar{\partial}$--invariance of (27) follows from the formal
$\bar{\partial}$--invariance of (28).  The action of $\bar{\partial}$
on (28), using the hierarchy (14), leaves us with total derivative
forms $\sim \oint ds \partial_s (\hat{V}_1.u_2)$ which are non--
analytic and cannot be set to 0.  However the partial--integration
procedure which leads to the physical density (27) has as a
supplementary effect that all such forms actually disappear from the
computation of $\bar{\partial} w_4$.  In fact, the partial--
integration procedure defines $Tr \, {\cal L}^4$ as (27) in
precisely the right way to have $\bar{\partial} w_4 = 0$.

This supplementary result confirms in a very non--trivial way our
claim that $w_3 (z)$ in (20)
is indeed the spin 2 $w_{\infty}$--generator for the $2+1$ Toda field theory.

It is now interesting to try to obtain higher spin generators.
However, it must be emphasized that formal part--integration itself
is
not sufficient to obtain well--defined, analytic higher
currents $w_n (z) , n\geq 5 ,$ from the ill--defined $Tr
\, {\cal L}^n$.  For instance, part--integrating $Tr \, {\cal L}^5$
ultimately leaves a ``residual" term:
$$\int ds \cdot u_2 (u_0 ^2 + 2u_1 )\eqno\eq$$
which cannot be made analytic by simply further part--integrating.
The identification of $w_n (z)$ with ``re--expressed"
$Tr \, {\cal L}^n$ breaks down at this level probably because the formal
object $Tr \, {\cal L}^n$ deviates too much from regular analytic
behaviour in $s$ to be of any meaning. It
makes it necessary
to devise more sophisticated ways of getting the analytical $w_n (z)$
if one wishes to avoid the cumbersome direct computation of higher
generators by Poisson brackets.  Clearly one should first compute
directly $w_5 (z)$ from $w_3 (z)$ and $w_4 (z)$ in order to have an
insight of such mechanisms.  We shall postpone this derivation
for further studies.

Finally we expect that this construction will lead, in the quantum
case, to operators $w_n (z)$ obeying nevertheless a classical--type
$w_{\infty}$--algebra. It is known [23] that the
limit $N\rightarrow \infty$ of the quantum
Gelfand-Dikii algebra is, by power--counting arguments, purely the
classical $w_{\infty}$--algebra plus a central charge in the Virasoro
sector.  Since the objects $w_n$ which we have constructed are
identified with the \underbar{continuous} large-$N$ limit of the
component fields obeying the Gelfand-Dikii algebra, their quantum
algebra must be $w_{\infty}$.

\noindent{\bf \underbar{Acknowledgements:}}\quad   I  wish to thank O.
Babelon, A. Jevicki and K. Takasaki for fruitful discussions and comments.

\noindent{\bf \underbar{References}}

\pointbegin
V. Fateev, A.B. Zamolodchikov, {\it Nucl. Phys.} {\bf B280}, 644
(1987).\nextline
V. A. Fateev, I. Lykyanov, {\it Int. Journ. Mod. Phys.} {\bf A7} 853 (1991).
\point
I. Bakas, {\it Comm. Mat. Phys.} {\bf 134}, 487 (1990).
\point
I. Bakas, E. Kiritsis, {\it Int. Journ. Mod. Phys.}, {\bf A6}, 2871
(1991).
\point C. Pope, L.Romans, X. Shen, {\it Phys. Lett.} {\bf B 236} , 173
(1990).
\point
M. Fukuma, H. Kawai, R. Nakayama, {\it Comm. Mat. Phys.} {\bf 143},
371 (1992).
\point
I. Bakas, E. Kiritsis, {\it Nucl. Phys.} {\bf B 343}, l65 (1990); X.
Shen, X. J. Wang, Preprint CPT-TAMU-59/91 (1991).
\point
V. G. Drinfel'd, V. V. Sokolov, {\it Journ. Sov. Mat.} {\bf 30}, 1975
(1984);\nextline
O. Babelon, Lectures XX Intern. G.I.F.T. Seminar (1989); PAR LPTHE 89-
29; {\it Nucl. Phys. B} (Proc. Suppl. 18A) (1990), 1.
\point
I. M. Gel'fand, L. A. Dikii, {\it Russ. Math. Surv.} {\bf 30}, 77
(1975).
\point
M. Bershadsky, H. Ooguri, {\it Comm. Mat. Phys.} {\bf 126}, 49 (1989).
\point
M. Saveliev, {\it Comm. Mat. Phys.} {\bf 121} , 283 (1989).
\point
K. Takasaki, T. Takebe, {\it Lett. Math. Phys.} {\bf 23}, 205 (1992).
\point
Q. Han Park, {\it Nucl. Phys.}  {\bf B 333}, 267 (1990).
\point
Q. Han Park, {\it Phys. Lett.} {\bf B236} ,423 (1990).
\point
A. G. Reyman, M. Semenov-Tjan-Shansky, {\it Inv. Mat.} {\bf 54}, 81
(1979).
\point
J. Avan, M. Bellon, {\it Phys. Lett.} {\bf B 213} , 459 (1988).
\point
J. Avan, A. Jevicki, in preparation.
\point
J.Avan, A. Jevicki, {\it Mod. Phys. Lett.} {\bf A7}, 357 (1992).
\point
I. M. Krichever, {\it Comm. Math. Phys.} {\bf 143}, 415 (1992).
\point
L. J. Romans, {\it Nucl. Phys.} {\bf B352}, 829 (1991).
\point
T. Takebe, {\it Comm. Math. Phys.} {\bf 129}, 181 (1990).
\point
D. Olive, N. Turok, {\it Nucl. Phys.} {\bf B265}, 469 (1986).
\point
A. V. Mikhailov, M. A. Olshanetsky, A. M. Perelomov, {\it Comm. Mat.
Phys.} {\bf 79}, 473 (1991).
\point
A. Bilal, {\it Phys. Lett.} {\bf B227}, 406 (1989).
\point
C. Pope, K. Stelle, {\it Phys. Lett.} {\bf B225}, 257 (1989).

\bye